\documentclass[apj,numberedappendix,twocolappendix,appendixfloats]{emulateapj}

\usepackage{rotating}
\usepackage{epsfig}

\usepackage[fleqn]{amsmath}

\newcommand{\lt}{\ifmmode\,<\,\else \,$<$\,\fi}
\newcommand{\kms}{\ifmmode\,{\rm km}\,{\rm s}^{-1}\else km$\,$s$^{-1}$\fi}

\newcommand{\magarc}{\ifmmode {{{{\rm mag}~{\rm arcsec}}^{-2}}}
             \else {{{mag}$~${arcsec}$^{-2}$}}
             \fi}
\newcommand{\hunit}{km~s$^{-1}$~Mpc$^{-1}$}
\newcommand{\hub}{H$_{\hbox{\scriptsize 0}}$}

\newcommand{\Lya}{Ly$\alpha$}
\newcommand{\oh}{$12+\log{\mathrm{O/H}}$}
\newcommand{\hst}{{\it HST}}
\newcommand{\clyi}{J0717}

\newcommand{\funit}{\mathrm{ergs}~\mathrm{s}^{-1}~\mathrm{cm}^{-2}}
\newcommand{\Msun}{\mathrm{M}_{\sun}}
\newcommand{\Msunyr}{\mathrm{M}_{\sun}~\mathrm{yr}^{-1}}

\newcommand{\Hii}{H~{\sc ii}}
\newcommand{\Oii}{[O~{\sc ii}]}
\newcommand{\Oiii}{[O~{\sc iii}]}
\newcommand{\Neiii}{[Ne~{\sc iii}]}
\newcommand{\Ha}{H$\alpha$}
\newcommand{\Hb}{H$\beta$}
\newcommand{\Hg}{H$\gamma$}
\newcommand{\Nii}{[N~{\sc ii}]}
\newcommand{\Sii}{[S~{\sc ii}]}

\usepackage{bbm}

\shorttitle{GLASS metallicity gradients}

\begin{document}

\title{The Grism Lens-Amplified Survey from Space (GLASS). II. Gas-phase metallicity and radial gradients in an interacting system at $z \simeq 2$}


\author{
T.~Jones\altaffilmark{1},
X.~Wang\altaffilmark{1},
K.~B.~Schmidt\altaffilmark{1},
T.~Treu\altaffilmark{1,2},
G.~B.~Brammer\altaffilmark{3},
M.~Brada\v{c}\altaffilmark{4},
A.~Dressler\altaffilmark{5},
A.~L.~Henry\altaffilmark{6},
M.~A.~Malkan\altaffilmark{2},
L.~Pentericci\altaffilmark{7},
M.~Trenti$^{8}$
}

\altaffiltext{1} {Department of Physics, University of California, Santa Barbara, CA  93106, USA}
\altaffiltext{2} {Department of Physics and Astronomy, UCLA, Los
Angeles, CA, USA 90095-1547}
\altaffiltext{3} {Space Telescope Science Institute, 3700 San Martin Drive, Baltimore, MD, 21218, USA}
\altaffiltext{4} {Department of Physics, University of California, Davis, CA, 95616, USA}
\altaffiltext{5} {The Observatories of the Carnegie Institution for Science, 813 Santa Barbara St., Pasadena, CA 91101, USA}
\altaffiltext{6} {Astrophysics Science Division, Goddard Space Flight Center, Code 665, Greenbelt, MD 20771, USA}
\altaffiltext{7} {INAF - Osservatorio Astronomico di Roma Via Frascati 33 - 00040 Monte Porzio Catone, Italy}
\altaffiltext{8} {Institute of Astronomy and Kavli Institute for Cosmology, University of Cambridge, Madingley Road, Cambridge, CB3 0HA, UK}


\begin{abstract}

We present spatially resolved gas-phase metallicity for a system of three galaxies at $z=1.85$ detected in the {\it Grism Lensed-Amplified Survey from Space} (GLASS). The combination of \hst's diffraction limit and strong gravitational lensing by the cluster MACS J0717+3745 results in a spatial resolution of $\simeq 200-300$ pc, enabling good spatial sampling despite the intrinsically small galaxy sizes. The galaxies in this system are separated by $\simeq 50-200$ kpc in projection and are likely in an early stage of interaction, evidenced by relatively high specific star formation rates. Their gas-phase metallicities are consistent with larger samples at similar redshift, star formation rate, and stellar mass. We obtain a precise measurement of the metallicity gradient for one galaxy and find a shallow slope compared to isolated galaxies at high redshift, consistent with a flattening of the gradient due to gravitational interaction.
An alternative explanation for the shallow metallicity gradient and elevated star formation rate is rapid recycling of metal-enriched gas, but we find no evidence for enhanced gas-phase metallicities which should result from this effect. Notably, the measured stellar masses $\log{\rm{M}_*/\Msun} = 7.2-9.1$ probe to an order of magnitude below previous mass-metallicity studies at this redshift. The lowest mass galaxy has properties similar to those expected for Fornax at this redshift, indicating that GLASS is able to directly study the progenitors of local group dwarf galaxies on spatially resolved scales. Larger samples from the full GLASS survey will be ideal for studying the effects of feedback, and the time evolution of metallicity gradients. These initial results demonstrate the utility of \hst\ spectroscopy combined with gravitational lensing for characterizing resolved physical properties of galaxies at high redshift.

\end{abstract}

\keywords{galaxies: evolution --- galaxies: ISM}

\section{Introduction}\label{sec:intro}

Galaxy gas-phase metallicities encode information about the history of gas accretion, star formation, and gaseous outflows. Measurements of metallicity combined with accumulated stellar mass and gas content therefore provide stringent constraints on the baryonic processes relevant to galaxy formation. Current observations show that essentially all galaxies have low gas-phase metallicities compared to estimated chemical yields from star formation, with greater discrepancy at lower stellar masses \citep[the mass-metallicity relation, e.g.,][]{Tremonti2004, Lequeux1979}. This is commonly attributed to outflows of metal-enriched gas driven by intense star formation, supported by observations that such feedback is ubiquitous among galaxies with high densities of star formation \citep{Heckman2001, Newman2012}.
The {\it spatial} distribution of metallicity within a galaxy further constrains the baryonic assembly history, in particular the effects of outflows and dynamical evolution. Star-forming disk galaxies in the local universe have negative radial metallicity gradients (higher metallicity in the central regions; see \citealt{Sanchez2014} and \citealt{Vila-Costas1992} for large samples). This has been explained by models of chemical evolution in which galaxies grow {\it inside-out} such that outer regions have younger characteristic ages \citep[e.g.,][]{Nelson2012}. Gradients typically flatten at large radii indicating efficient mixing in those regions, possibly due to secular dynamical evolution or recycling of metal-enriched gas which was previously ejected in outflows \citep[so-called "galactic fountains"; e.g.][]{Werk2011, Bresolin2012}. Notably, galaxies which are undergoing mergers or strong interactions have significantly shallower metallicity gradients than isolated disks, which is understood numerically in terms of radial gas flows induced by tidal forces \citep[most notably metal-poor inflows;][]{Rupke2010a, Rupke2010b, Sanchez2014, Rich2012}.

This work is concerned with using time evolution of metallicity gradients as a probe of galaxy formation. We seek to measure how gradients evolve, and to understand what processes drive that evolution. A variety of predictions have been made based on numerical models of inside-out growth, ranging from steeper to flatter to inverted (positive) gradients at higher redshifts \citep[e.g.,][]{Prantzos2000, Chiappini2001, Magrini2007, Fu2009}. Cosmological simulations have recently begun to address this issue, with several groups arguing that gradient evolution depends strongly on star formation feedback: stronger feedback is predicted to cause flatter gradients due to rapid gas recycling \citep{Pilkington2012, Gibson2013, Angles-Alcazar2014}. Meanwhile the observational results at high redshift have grown in number but comprise a variety of conclusions for the evolution of gradients. Data reaching $\lesssim1$ kpc resolution with adaptive optics (and up to $\sim$100 pc in the case of lensed galaxies) have revealed negative radial gradients with slopes often steeper than local descendants, suggesting that gradients flatten as galaxies grow with time \citep{Jones2010,Jones2013,Yuan2011,Swinbank2012}. Larger surveys with seeing-limited resolution of $\sim$5 kpc have reported mostly flat gradients with a significant fraction of positive slopes \citep{Cresci2010,Queyrel2012,Stott2014,Troncoso2014}. These have been interpreted as evidence for predicted "cold flows" of pristine gas \citep[e.g.,][]{Dekel2009,Keres2009,Faucher-Giguere2011}, but these results are in contrast with high-resolution measurements.

The discrepancy in metallicity gradient measurements at high redshift must be addressed in order to reliably understand the role of gas and metal transport in galaxy evolution. High spatial resolution is clearly an advantage; \citet{Yuan2013} demonstrate how degraded resolution can dramatically bias gradient measurements. For typical-size galaxies at high redshift, $\lesssim1$ kpc resolution is essential to sample the scale radius and reliably measure gradient slopes. The use of single strong-line metallicity indicators such as \Nii$/$\Ha\ \citep[e.g.,][]{Pettini2004} is another potential problem. Multiple emission line diagnostics are necessary  to confirm variations in metallicity \citep[e.g.,][]{Jones2013} and to distinguish \Hii\ regions from AGN or shock excitation, which in many cases are non-negligible \citep{Wright2010,Yuan2012,Newman2014} and would bias the inferred gradients. In essence we require larger samples with high spatial resolution and a reliable suite of metallicity and excitation diagnostics in order to resolve the discrepancy in existing measurements. Here we present initial results from the {\it Grism Lensed-Amplified Survey from Space} (GLASS) in which we measure metallicities at $z\simeq2$ based on 5 nebular emission lines with resolution as fine as 200 pc. As with \citet{Jones2013} this is enabled by a combination of strong gravitational lensing, broad wavelength coverage, and diffraction-limited data. These results demonstrate the potential of the full GLASS survey to obtain dozens of such measurements at $1.3\lesssim z \lesssim2.3$ and thus reliably determine the average evolution of metallicity gradients over the past 11 Gyr.

Throughout this paper we adopt a flat $\Lambda$CDM cosmology with $\Omega_{\Lambda}=0.7$, $\Omega_{\rm M}=0.3$, and \hub$=70$ \hunit. At $z=1.85$, 3.5 Gyr has elapsed since the big bang and 1 arcsecond corresponds to 8.4 kpc (c.f. 3.5 Gyr and 8.7 kpc/arcsecond for the \citealt{Planck2013} cosmology). All stellar masses (M$_*$) and star formation rates (SFR) correspond to a \cite{Chabrier2003} initial mass function. Unless stated otherwise, "metallicity" refers to the gas-phase oxygen abundance.

\section{A triple galaxy system at $z=1.855$}

The first GLASS observations of MACS J0717+3745 (hereafter \clyi) revealed three different strongly lensed sources at the same redshift $z=1.855\pm0.01$ (arc systems 3, 4, and 14 in \citealt{Schmidt2014}). The grism observations cover three magnified images of each galaxy (e.g., arcs 3.1, 3.2, and 3.3 are multiple images of the same source) and the details of each image are given in \cite{Schmidt2014}. The gravitational lensing model discussed in Section~\ref{sec:model} indicates projected galaxy separations of 50 kpc for pairs 4--14, 150 kpc for 3--14, and 200 kpc for 3--4. Although the grism redshifts cannot constrain their line-of-sight separation to better than a few Mpc, Keck/LRIS spectra of arcs 3.2 and 14.1 \citep{Limousin2012} and a Keck/MOSFIRE spectrum of arc 4.1 \citep{Schmidt2014} indicate consistent redshifts, corresponding to $\lesssim1$ Mpc in the Hubble flow. Therefore all three galaxies appear to be physically associated. We argue in Section~\ref{sec:results} that arcs 4 and 14 are likely affected by significant forces from gravitational interaction on the basis of their physical properties, however we caution that their true 3-D separation is uncertain.
The fortuitous inclusion of these systems within \hst's field of view, and the availability of several strong emission lines in the grism spectra, provide a laboratory for studying how gas and heavy elements cycle within galaxies and their surrounding medium in an overdense environment.

\section{Observations and data analysis}\label{sec:data}

Details of the GLASS survey design and grism spectroscopy are given in \citet{Schmidt2014}. Briefly, J0717 was observed at two different position angles separated by approximately 100 degrees. The total exposure time is 10 orbits in the G102 grism and 4 orbits in the G141 grism. Data from each position angle are interlaced to produce 2D grism spectra and direct images with a scale of 0\farcs065 per pixel, Nyquist sampling the point spread function. Direct images are used to estimate continuum spectra by dispersing each pixel from individual objects detected in {\sc SExtractor} \citep{Bertin1996} segmentation maps; these are used to guide the extraction of individual object spectra and to subtract an estimate of the contamination from nearby objects.

\begin{figure}
\includegraphics[width=0.48\textwidth]{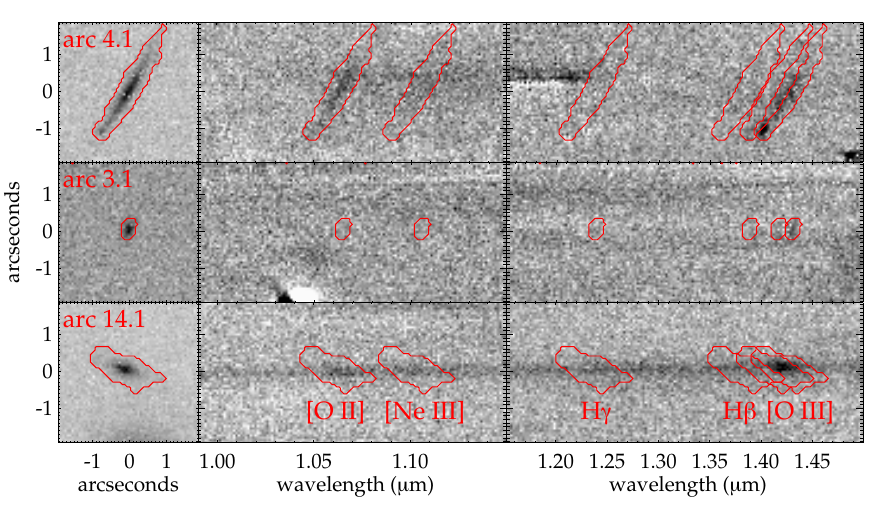}
\caption{
\label{fig:spec2d}
Grism spectra for the arcs of interest show multiple spatially extended emission lines which can be used to derive metallicity maps. From left to right, each row shows the F140W direct image, G102 grism spectrum, and G141 grism spectrum. In all cases the estimated contamination has been subtracted from the spectra. Strong source continuum is apparent for arc 14.1. Red contours show the object segmentation maps derived using the direct images, and mapped to emission lines of interest in the grism spectra using the redshift of each source (left to right: \Oii\ $\lambda\lambda$3727, \Neiii\ $\lambda$3869, \Hg, \Hb, \Oiii\ $\lambda$4959, \Oiii\ $\lambda$5007). The redshift is $z=1.855$ for all cases shown here.
}
\end{figure}

\begin{figure}
\includegraphics[width=0.5\textwidth]{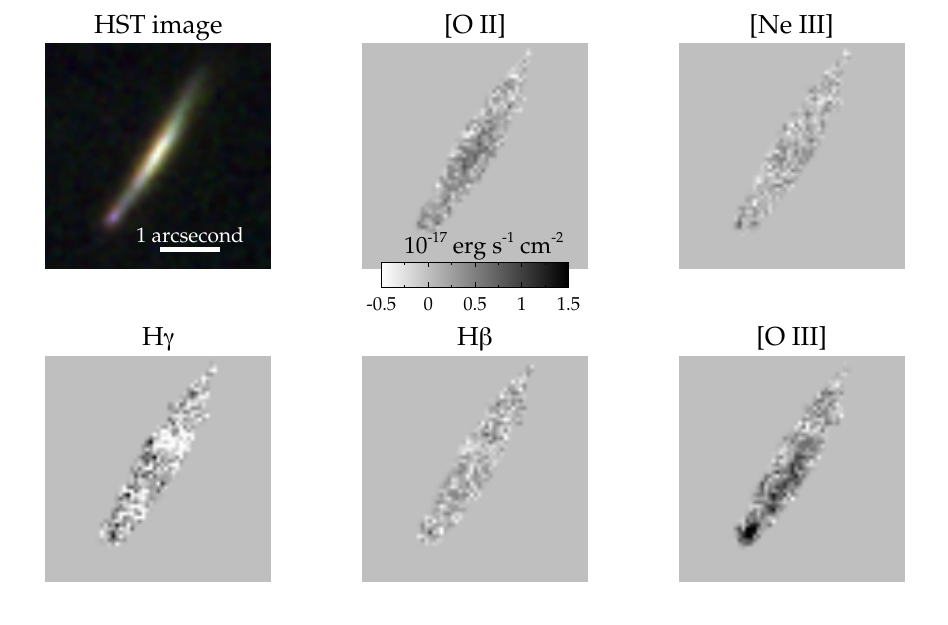}
\caption{
\label{fig:spec2d_arc4}
\hst\ image (top right; RGB: WFC3/F160W, WFC3/F110W, ACS/F814W) and emission line maps of arc 4.1. The typical flux uncertainty in each pixel is $2 \times 10^{-18}$ $\funit$ (1$\sigma$).
}
\end{figure}

The critical analysis method for this work is extraction of emission line maps. For each object of interest, we use the direct image to construct a segmentation map and shift this to the expected wavelength of strong emission lines using spectroscopic redshifts from the grism data (Figure~\ref{fig:spec2d}). We extract two-dimensional maps of emission line flux and uncertainty for the region defined by the segmentation map, and correct for residual background structure by subtracting a corresponding map at a nearby line-free wavelength (e.g., a rest-frame 4750 \AA\ map is typically used to correct \Hb\ for imperfect continuum and contamination estimates). Data affected by strong contamination are masked and treated as missing. We align the emission line maps extracted from both position angles using bilinear interpolation, and combine them with an inverse-variance weighting. Example emission line maps are shown for arc 4.1 in Figure~\ref{fig:spec2d_arc4}.

The \Hb\ and \Oiii\ emission lines require special treatment in cases where  the spatial extent is sufficiently large that these lines overlap. In all cases there is a region of pure \Oiii\ $\lambda$5007 emission at the rightmost edge of the blend (of $\simeq 6$ pixels or $0\farcs4$ in the dispersion direction for $z=1.855$). We use the intrinsic flux ratio $f_{5007}/f_{4959} = 3$ to subtract the corresponding \Oiii\ $\lambda$4959 flux, which leaves another region of pure \Oiii\ $\lambda$5007 emission. We iterate this process to completely de-blend the \Oiii\ doublet, resulting in separate $\lambda$4959 and $\lambda$5007 emission line maps. Additionally we de-blend \Oiii\ $\lambda$4959 and \Hb, although we cannot completely de-blend the lines in cases where \Oiii\ $\lambda$5007 and \Hb\ overlap. For $z=1.855$ this occurs whenever the source size is $\geq 1\farcs4$ in the dispersion direction (for example, the spectrum of arc 14.1 shown in Figure~\ref{fig:spec2d}). In such cases we treat the region where \Oiii\ $\lambda$5007 and \Hb\ overlap as missing data.

\section{Gravitational lensing model}\label{sec:model}

An accurate gravitational lensing model is essential for reconstructing the source plane morphology of lensed galaxies, and for combining the information from the multiple images. As one of the Frontier Fields, the gravitational lensing potential of \clyi\ has been modeled by several groups using a variety of techniques, and the results are publicly available\footnote{http://www.stsci.edu/hst/campaigns/frontier-fields/}. We compared the results of all models for which we could derive deflection angle maps at the redshift of interest (those of Brada{\v c}, CATS, Sharon, and Zitrin) to assess which is best suited to the purposes of this work. We note that those models are aimed at producing a global description of the cluster and therefore their accuracy in the vicinity of the lensed images of interest is expected to vary significantly between them. The Sharon version 2 model \citep{Johnson2014} produced the best results for our images, yielding the most precise inversion. The precision of the inversion was evaluated by comparing for each set of multiple images the promixity of the inferred source position, and the agreement beetween source plane flux and morphology.  However, even the best global model produced significant residual differences between the reconstructed sources, requiring an additional step, as described in the next paragraph.

In order to take full advantage of the multiple images of each arc system, we have developed a novel technique to align all images in the source plane. This allows us to reduce the lensing-related uncertainties and combine the data from multiple images to increase the signal-to-noise ratio of emission line maps. The full details of our methodology will be presented by Wang et al. (2014, in prep); here we give only a brief overview. Essentially, we are considering the global cluster model as an approximate first solution and we are seeking corrections to the potential to improve the reconstruction. We assume that the corrections are small and can thus be described by a local correction to the lensing potential  up to the first two orders of derivatives.  The first order term consists of a correction to the deflection angle for each image. The second order term yields corrections to the shear and convergence. The procedure has thus five free parameters per image. The optimal parameters are found by requiring the source plane reconstructions of each set of multiple images to be as similar to each other as possible. We note that a direct byproduct of this formalism is a correction to the magnification of each image.

\begin{figure}
\includegraphics[width=0.5\textwidth]{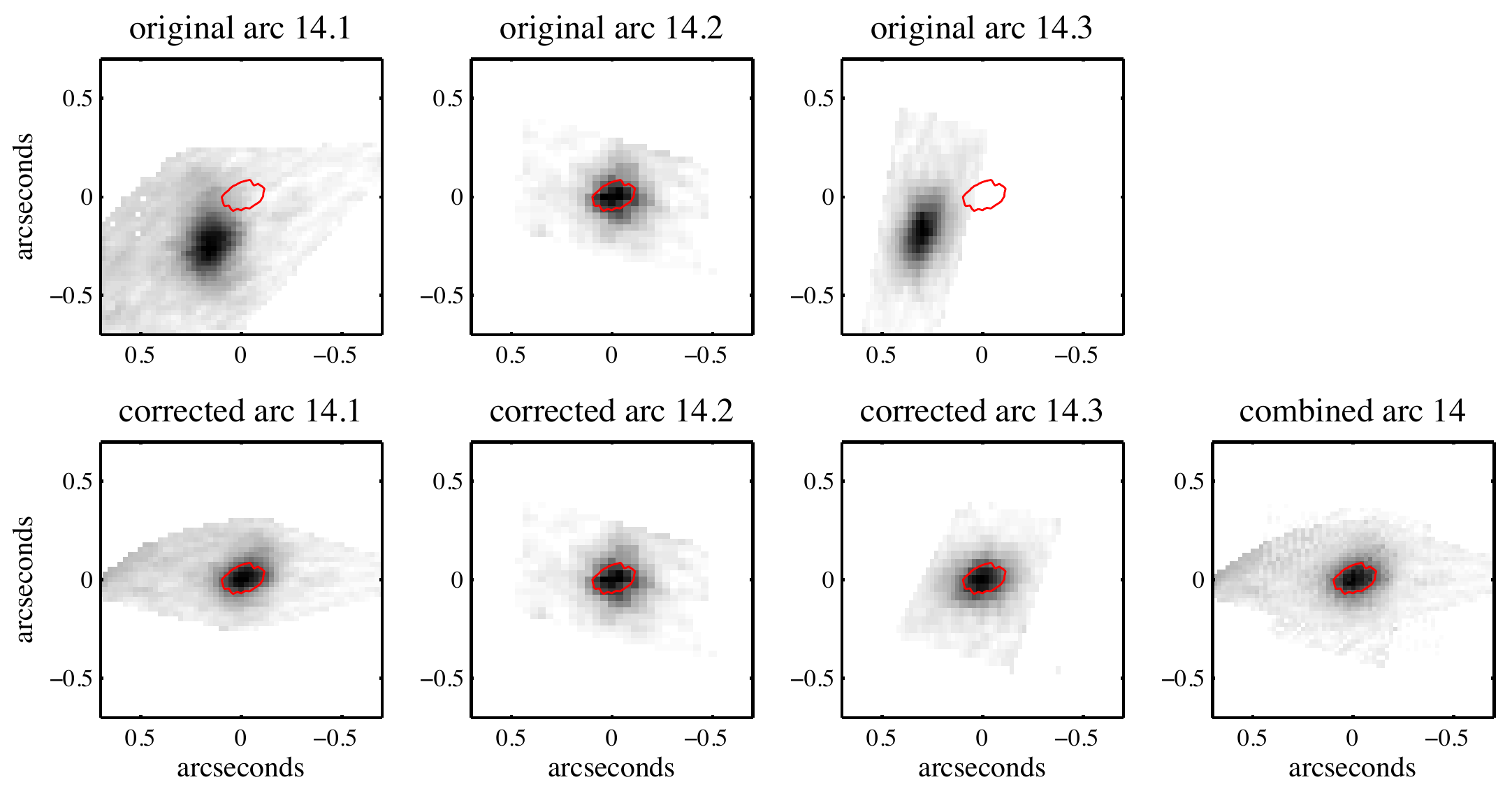}
\caption{
\label{fig:lenscorr}
Demonstration of the lens model corrections. The panels in the top row
are source plane reconstructions given by the original Sharon version
2 model, whereas those in the bottom row show the improved
reconstructions after our correction is applied. Note that the
original and corrected arc 14.2 is identical since it is used as a
reference. In all panels, the gray scale represents the surface
brightness contrast and the red contour shows an isophotal radius
measured for the combined arc. Here we show the process for arc 14 as
an illustration. The corrections for arcs 3 and 4 give similar results.}
\end{figure}

We have applied this method to the arc systems 3, 4, and 14, using the least distorted (i.e., least magnified) image of each 
system as a reference. Figure~\ref{fig:lenscorr} shows multiple images of arc 14 reconstructed with and without the lens model 
corrections to illustrate the advantage. The source plane positions are offset by $\simeq 0\farcs3 \simeq 3$ kpc using the 
original model, while the model corrections produce consistent positions and morphologies for all three reconstructed images. The 
second order correction is thus sufficient for our purposes. In all cases the correction is relatively small, amounting to $<5$\% 
of the total lensing deflection angle.

In the following analysis we combine emission line maps from multiple images in the source plane in order to increase the measurement precision. However, including less highly magnified images provides a marginal improvement at the cost of degraded spatial resolution. We therefore use arcs 3.1$+$3.2, arc 4.1 only, and arcs 14.1$+$14.3 to optimize resolution and sensitivity. Final results are consistent with measurements from individual images, as well as with results derived from the original lens model. Combining multiple images improves the precision in metallicity gradients by an impressive factor of $\simeq2\times$ for arcs 3 and 14 by enabling finer spatial sampling and detection of more extended low surface brightness emission.
We derive a conservative uncertainty of 13\% RMS in the magnification
factors (prior to second order correction) by comparing measured flux ratios of multiple images with model-predicted magnification ratios. This error is not propagated in the following analysis, however it is small compared to other sources of uncertainty and has no effect on the results.

\begin{figure}
\includegraphics[width=.3\columnwidth]{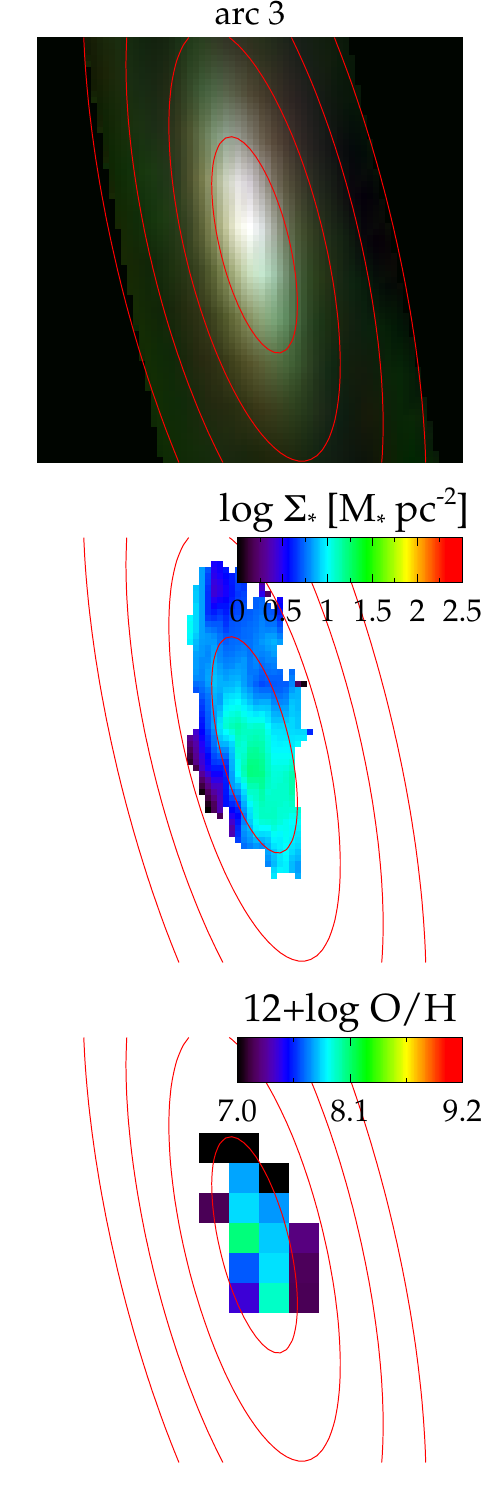}
\hspace{0.03\columnwidth}
\includegraphics[width=.3\columnwidth]{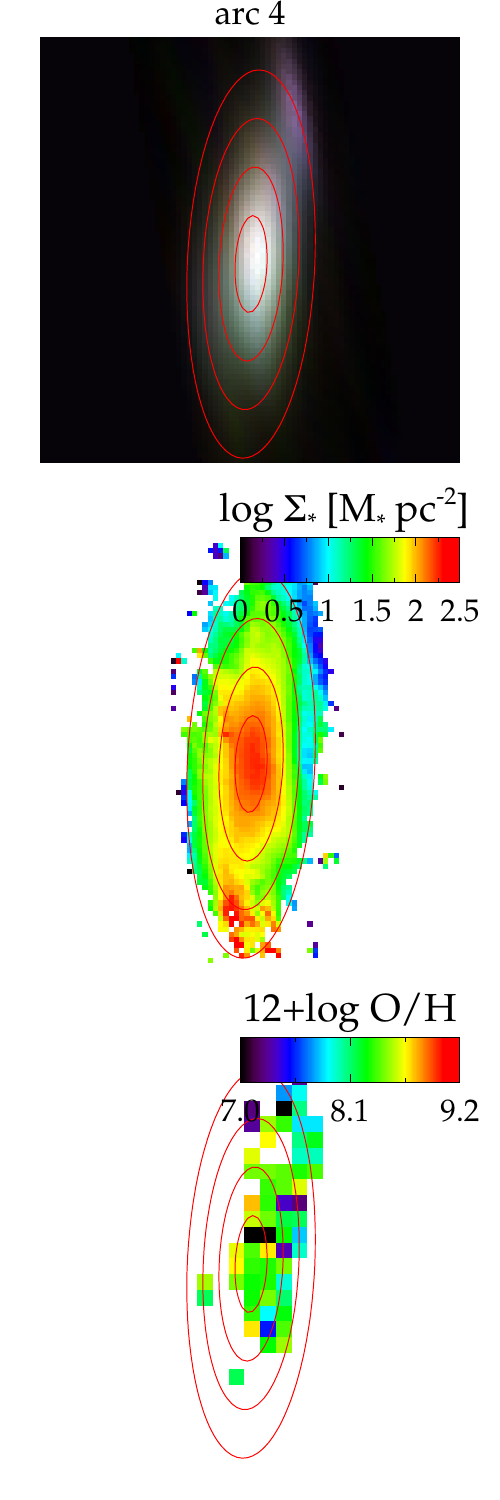}
\hspace{0.03\columnwidth}
\includegraphics[width=.3\columnwidth]{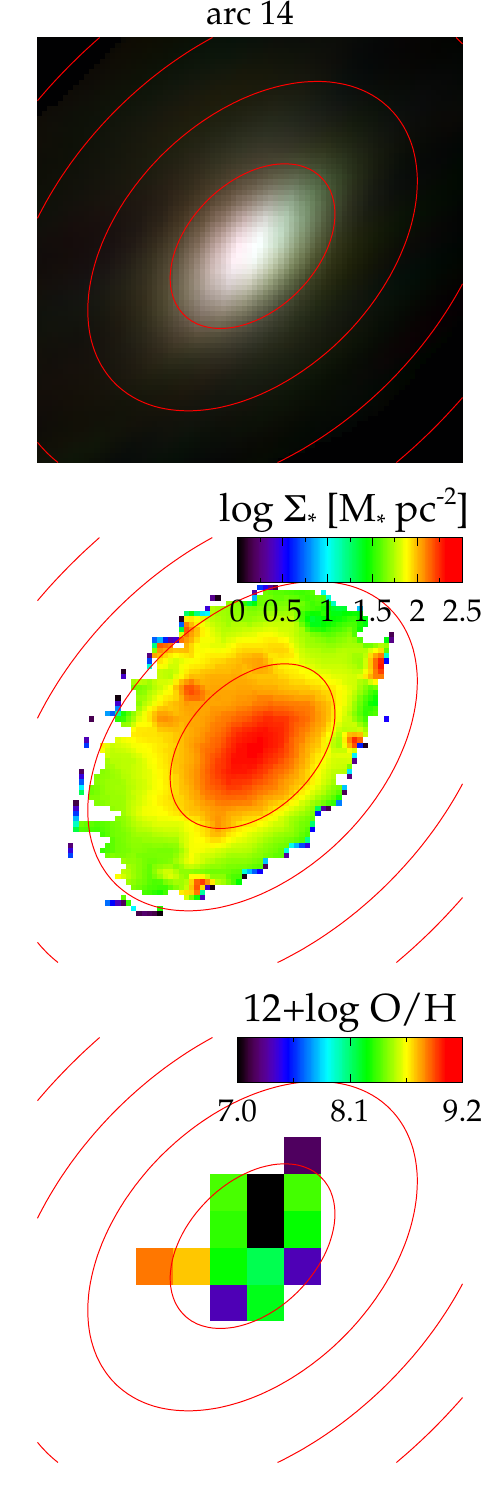}
\caption{
\label{fig:sourceplane}
Source plane morphologies of arcs 3, 4, and 14 (from left to right). Each panel shows contours of constant de-projected galactocentric radii at intervals of 1 kpc, derived in Section~\ref{sec:properties}.
{\it Top row:} \hst\ image (RGB: WFC3/F160W, WFC3/F110W, ACS/F814W).
{\it Middle:} stellar mass maps derived from spatially resolved \hst\ photometry. All galaxies have smooth, centrally peaked stellar mass profiles with no significant secondary peaks.
{\it Bottom:} gas-phase metallicity maps.
}
\end{figure}

\section{Physical properties}\label{sec:properties}

\subsection{Stellar mass}\label{sec:mstar}

We use the stellar population synthesis code FAST version 1.0 \citep{Kriek2009} to fit the resolved spectral energy density (SED) of each galaxy of interest. For $z=1.855$, the rest-frame UV through optical SEDs are well constrained by broad-band \hst\ photometry taken as part of the CLASH survey \citep{Postman2012}. We utilize the ACS/F435W, ACS/F606W, ACS/F814W, and WFC3/F125W filters which sample nearly the full rest-frame wavelength range from 1250--4900 \AA. This set of filters is chosen to provide the widest possible wavelength coverage while avoiding contamination from the strong emission lines \Lya, \Oii, and \Oiii\ which can significantly affect the broad-band photometry and derived stellar population properties. In particular, \Oiii\ emission accounts for $\sim$25\% of the total WFC3/F160W flux (an increase of 0.3 magnitudes) and \Oii\ significantly affects the WFC3/F105W flux for the galaxies discussed here.

Our methodology is as follows. We first align all images with the GLASS F105W direct image and smooth to a common point spread function of 0\farcs2 FWHM. The broad-band fluxes in each pixel are fit with a \citet{Bruzual2003} stellar population library, \cite{Chabrier2003} initial mass function, Milky Way dust attenuation law, stellar ages between 5 Myr and the age of the universe at the galaxy's redshift, and an exponentially declining star formation history with $\tau = 10^{7}-10^{10}$ yr. The quantity of greatest interest is the derived stellar mass surface density, which is the most robust parameter. Other stellar population parameters (SFR, age, extinction, etc.) are obtained simultaneously albeit with larger uncertainty. We have repeated the SED analysis using additional broad-band filters corrected for emission line contamination using the flux maps described in Section~\ref{sec:data}, verifying that this produces consistent results. Total stellar masses derived from integrated photometry and corrected for magnification are listed in Table~\ref{tab:arcs}.

\subsection{Morphology}\label{sec:morphology}

Morphological information is critical for measuring accurate metallicity gradients, as the radial coordinate depends on a galaxy's central position and inclination. Ideally the dynamical center, major axis orientation, and inclination would be constrained from kinematics as has been done for previous work \citep[e.g.,][]{Jones2013}, but we lack kinematic data. Instead we derive estimates of these quantities by assuming that the stellar mass surface density derived in Section~\ref{sec:mstar}, $\Sigma_*$, is elliptically symmetric. We reconstruct $\Sigma_*$ using the lens model (Section~\ref{sec:model}) and fit for the centroid, orientation, and axis ratio. Only the most highly magnified image in each arc system is used in order to maximize the spatial resolution. In the following analysis we adopt the galaxy center and inclination such that contours of $\Sigma_*$ trace contours of constant de-projected radius.

Source plane $\Sigma_*$ distributions and best-fit ellipses are shown in Figure~\ref{fig:sourceplane}. All galaxies exhibit smooth, centrally peaked stellar mass profiles with contours that are fit by ellipsoids. Within the stellar mass uncertainty, we find no evidence for ongoing late-stage major mergers which could manifest as a secondary peak in the stellar mass density.

\subsection{Metallicity and nebular extinction}\label{sec:z}

We use the strong line ratio calibrations presented by \cite{Maiolino2008} to estimate gas-phase oxygen abundance (expressed as \oh) and nebular extinction A(V) from measured \Oii, \Neiii, \Hg, \Hb, and \Oiii\ fluxes. An advantage of these lines is that they directly trace the oxygen abundance, as opposed to \Nii-based diagnostics which may suffer from redshift-dependent systematic errors arising from evolution in the N/O ratio or other effects \citep[e.g.,][]{Shapley2014,Steidel2014}.
We use a \cite{Cardelli1989} extinction curve with R$_{\rm V} = 3.1$, noting that the choice of R$_{\rm V}$ has no significant effect on the derived metallicity (typically $<0.1$ dex for $2\leq {\rm R_V} \leq 5$).
To further constrain A(V), we impose
$\frac{\rm{H}\gamma}{\rm{H}\beta} = 0.47$
as expected for Case B recombination and typical \Hii\ region conditions \citep[e.g.,][]{Hummer1987}.
Metallicity and extinction are derived from a $\chi^2$ statistic constructed from all available diagnostics:
$$\chi^2 = \sum_R \left[ \frac{\Delta R}{\sigma(R)} \right]^2.$$
Here $\Delta R$ is the difference between predicted and observed de-reddened flux ratios at a given A(V) and \oh, and the uncertainty $\sigma(R)$ includes RMS scatter in each calibration added in quadrature with measurement uncertainty. The diagnostic ratios used are R$_{23}$ $=$ (\Oii+\Oiii)/Hb, \Oiii/\Oii, \Oiii/Hb, \Oii/Hb, \Neiii/\Oii, and \Hg/\Hb. We adopt the best fit as the values that minimize $\chi^2$, and 1$\sigma$ uncertainty from the extrema at which $\chi^2$ differs by 1 from the minimum value. This is slightly smaller than the formal uncertainty because of covariance between various metallicity diagnostics; we have checked that a more careful treatment does not affect the conclusions. We limit the solutions to A(V)~$=0-3$ and \oh~$=7.0-9.3$ \citep[equivalent to $0.02-4.1\times$ the solar abundance of \oh~$=8.69$;][]{Allende2001}. While the formal best-fit solutions are occasionally beyond this range, all such cases are poorly constrained and we view such extreme conditions as highly unlikely. Although the extinction is poorly constrained for most individual pixels, SED fitting (Section~\ref{sec:mstar}) and averaged \Hb$/$\Hg\ ratios indicate relatively low extinction toward both the stars and \Hii\ regions, with best-fit A(V)$=0-0.6$ and permitted A(V)$<1.1$ (1$\sigma$) in the dustiest case. Fortunately, uncertainty in extinction has little effect on the derived metallicity since many diagnostic line ratios are relatively insensitive to reddening.

Maps of the source plane gas-phase metallicity are shown in Figure~\ref{fig:sourceplane}. The emission line maps are binned to spatial scales of $\simeq300-500$ pc and the metallicity is fit for all such pixels where at least one emission line is detected at $\geq 5\sigma$ significance. All sources are well resolved with measurements extending over $\simeq3$ resolution elements for arc 3, and $\geq 6$ resolution elements for the others. The metallicity is typically sub-solar with most pixels having best-fit \oh\ in the range 7.5$-$8.5, corresponding to flux ratios $\frac{\rm [O~\textsc{iii}]}{\rm{H} \beta} > 2.5$ and $\frac{\rm [O~\textsc{iii}]}{\rm [O~\textsc{ii}]} > 1$. Figure~\ref{fig:sourceplane} also shows several pixels for which the best-fit metallicity is at the extreme ends of the permitted range; these usually correspond to noise artifacts and/or single-line detections and are therefore unreliable.

In addition to the spatially resolved analysis, we calculate galaxy-averaged metallicity and extinction by applying the same analysis to total emission line fluxes measured from integrated 1-D spectra. The results agree with the average of individual pixel values (Table~\ref{tab:arcs}), providing a valuable sanity check. The slightly lower average metallicity of individual pixels c.f. integrated flux is likely due to the requirement of a 5$\sigma$ detection of \Oiii\ (the brightest emission line), which induces a bias toward lower metallicity. Extinction-corrected Balmer line fluxes derived from this analysis are used to determine star formation rates in Section~\ref{sec:sfr}.

\subsection{Metallicity gradient}\label{sec:gradients}

\begin{figure*}
\includegraphics[width=.33\textwidth]{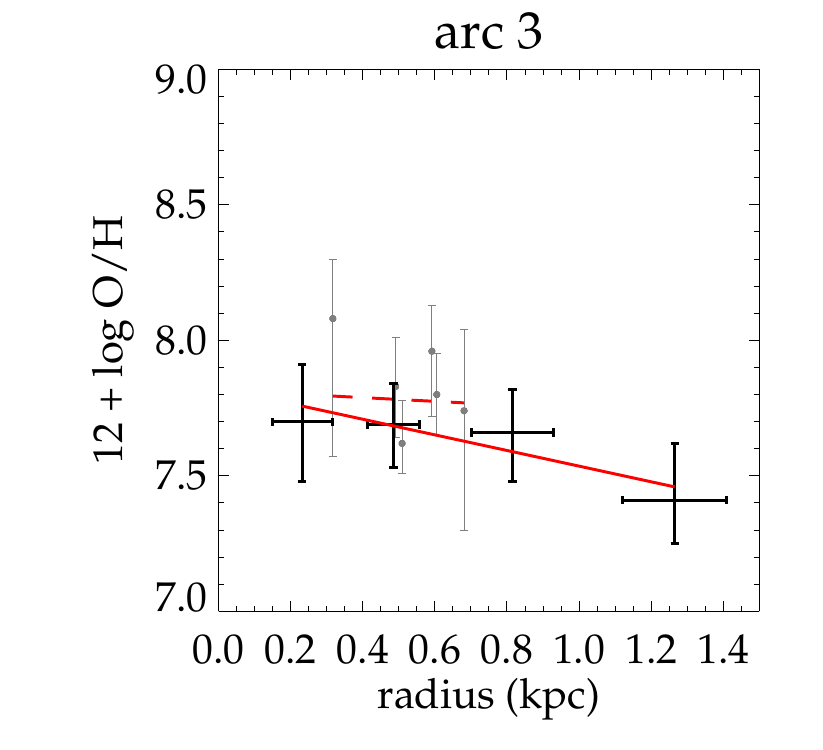}
\includegraphics[width=.33\textwidth]{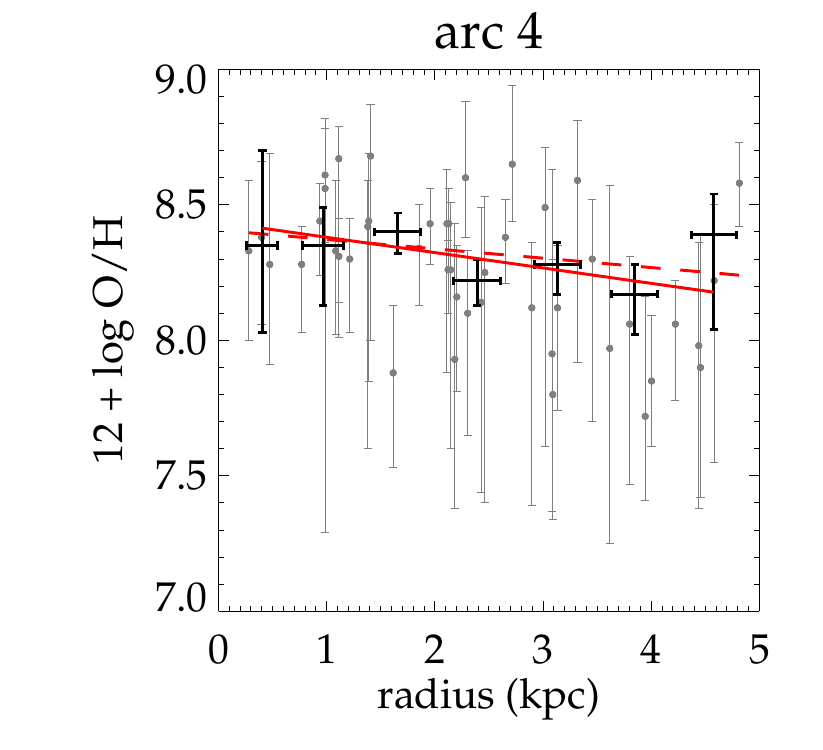}
\includegraphics[width=.33\textwidth]{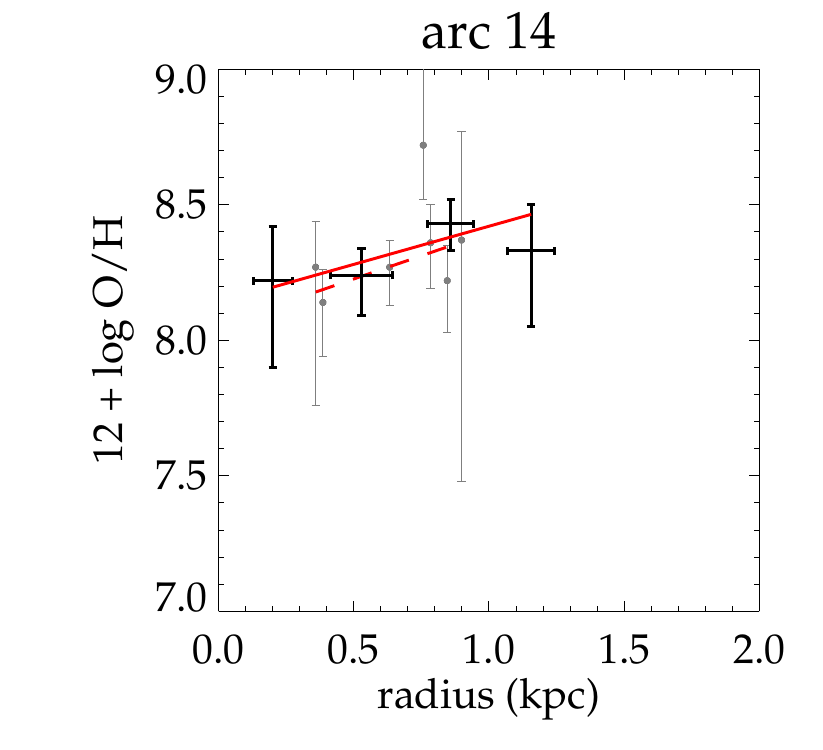}
\caption{
\label{fig:gradients}
Radial metallicity gradients in each galaxy. Grey points are measured from individual source plane pixels shown in Figure~\ref{fig:sourceplane}, and thick black points show the results for flux summed within radial annuli. Linear gradient fits to the individual pixels and radial bins are shown as dashed and solid lines, respectively. Both methods give consistent results.}
\end{figure*}

Gas-phase metallicity gradients are derived from the morphology and metallicity analyses described in Sections~\ref{sec:morphology} and \ref{sec:z}, respectively. In essence we measure the mean metallicity as a function of galactocentric radius in the source plane, utilizing knowledge of the major axis and inclination of each galaxy. We consider two approaches based on (1) individual pixels and (2) radial binning. For the first method, we compute the gradient $\frac{\Delta Z}{\Delta R}$ from a linear fit to the metallicity vs. de-projected radius of individual pixels (those shown in Figure~\ref{fig:sourceplane}):
\begin{equation}\label{eq:gradient}
$\oh$\, = Z_0 + \frac{\Delta Z}{\Delta R} R.
\end{equation}
This functional form is a good fit to the data and we are unable to distinguish more complex behavior given the measurement uncertainties. Pixels for which the allowed (1$\sigma$) range of \oh\ extends to $\leq7$ or $\geq9.3$ are excluded from the fit as these are generally unreliable and have large uncertainty ($\gtrsim 1$ dex), although including these data gives consistent results. For the second method we bin the emission line flux in annular apertures for each source, calculate metallicity from the total flux in each aperture, and compute the gradient using Equation~\ref{eq:gradient}. The range of radii for each annulus is chosen to provide a $\simeq10\sigma$ detection of \Oiii, which is uniformly the strongest emission line. We show the data and best-fit gradients in Figure~\ref{fig:gradients} and list the results in Table~\ref{tab:arcs}. Uncertainty in inclination and magnification are not included in these error estimates, but these effects are small ($\lesssim 30$\% combined) compared to uncertainty in the metallicities.

Both methods of calculating the metallicity gradient use the same diagnostics applied to the same data set and give consistent results. The pixelated approach is similar to local galaxy measurements which are based on individual \Hii\ regions \citep[e.g.,][]{Vila-Costas1992}, and the radial binning method is equivalent to most measurements reported for high redshift galaxies \citep{Yuan2011,Swinbank2012,Queyrel2012,Stott2014}. The primary differences are that the annular binning method is higher signal to noise ratio and is spatially complete, whereas low surface brightness regions are either noisy or excluded from the pixelated method. We therefore consider the annular binning method to be more robust.

\subsection{Star formation rate}\label{sec:sfr}

Star formation rates are derived from Balmer emission using standard methods. We convert the extinction-corrected total \Hb\ flux (Section~\ref{sec:z}) to SFR following \cite{Kennicutt1998}, divide by 1.7 to convert to a \cite{Chabrier2003} IMF, and divide by the lensing magnification to recover the intrinsic SFR of each galaxy. Results are given in Table~\ref{tab:arcs} including uncertainty from measurement noise and extinction. Additional uncertainties associated with the choice of $R_V$ and underlying stellar absorption (for which we make no correction) are negligible.

\subsection{AGN contamination}

In calculating metallicities and SFR we have assumed that all nebular emission arises in \Hii\ regions. The presence of shocks or AGN can bias these results even for very weak nuclear sources and we therefore check for any possible contamination. The global flux ratios of each galaxy, as well as the individual annular bins shown in Figure~\ref{fig:gradients}, are consistent with \Hii\ regions. However in all cases the line ratios fall in the region occupied by both \Hii\ and AGN excitation \citep[for example in the "blue" diagnostic diagram of \Oiii/Hb vs. \Oii/Hb, e.g.,][]{Lamareille2010}, and we lack a low-excitation line such as \Nii\ or \Sii\ which would permit more reliable classification \citep[e.g.,][]{Jones2013}. Publicly available Chandra/ACIS data\footnote{Available at http://cxc.cfa.harvard.edu/cda/} show no X-ray detection for any of the arcs, and all three arcs have moderate-resolution Keck spectra which show no sign of AGN features \citep{Schmidt2014,Limousin2012}. We note that the central resolution element of arc 14 shows marginal evidence ($\sim2\sigma$) of an elevated \Oiii$/$\Hb\ ratio as expected for AGN \citep[e.g.,][]{Wright2010,Newman2014}.
\cite{Trump2011} have associated this signal with a significant AGN fraction in stacks of galaxies of similar stellar mass and redshift whose individual X-ray luminosities are below the detection threshold, although it is not a reliable indicator of AGN in individual galaxies.
To summarize, we find no conclusive evidence of AGN but cannot rule out a possible low-level contribution to the emission line flux from a nuclear source. The expected effect of such activity would be to lower the derived central metallicity, with no significant effect on SFR or stellar masses.

\begin{deluxetable*}{lcccccccc}
\tablecolumns{8}
\tablewidth{0pt}
\tablecaption{Source properties\label{tab:arcs}}

\tablehead{
\colhead{ID} & \colhead{$z$} & \colhead{log M$_*$}      & \colhead{SFR}          & \colhead{\oh\tablenotemark{1}} & \colhead{\oh\tablenotemark{2}} & \colhead{$\Delta \rm{Z} / \Delta \rm{R}$\tablenotemark{3}} & \colhead{$\Delta \rm{Z} / \Delta \rm{R}$\tablenotemark{4}} \\
\colhead{}   & \colhead{}    & \colhead{[log $\Msun$]}  & \colhead{[$\Msunyr$]}  & \colhead{}                 & \colhead{}                       & \colhead{[dex kpc$^{-1}$]}      & \colhead{[dex kpc$^{-1}$]} \\
}
\medskip
\startdata
arc 3   &  1.855  &  7.2$\pm$0.4          &  1.8$^{+1.0}_{-0.8}$  &  7.56$\pm$0.37  &  7.75$^{+0.25}_{-0.19}$  &  -0.07$\pm$1.06  &  -0.29$\pm$0.25  \\
arc 4   &  1.855  &  9.1$^{+0.2}_{-0.1}$  &   29$^{+21}_{-14}$    &  8.20$\pm$0.34  &  8.35$^{+0.11}_{-0.13}$  &  -0.03$\pm$0.03  &  -0.05$\pm$0.05  \\
arc 14  &  1.855  &  8.8$^{+0.2}_{-0.1}$  &  3.3$^{+3.2}_{-0.9}$  &  8.11$\pm$0.31  &  8.30$^{+0.15}_{-0.23}$  &   0.34$\pm$0.41  &   0.28$\pm$0.29  \\
\enddata
\tablenotetext{1}{Average and RMS scatter of individual pixels}
\tablenotetext{2}{Best fit and 1$\sigma$ uncertainty derived from integrated spectrum}
\tablenotetext{3}{Derived from individual pixels}
\tablenotetext{4}{Derived from total flux in radial apertures}
\end{deluxetable*}

\section{Results and discussion}\label{sec:results}

\subsection{Global properties}

\begin{figure}
\includegraphics[width=\columnwidth]{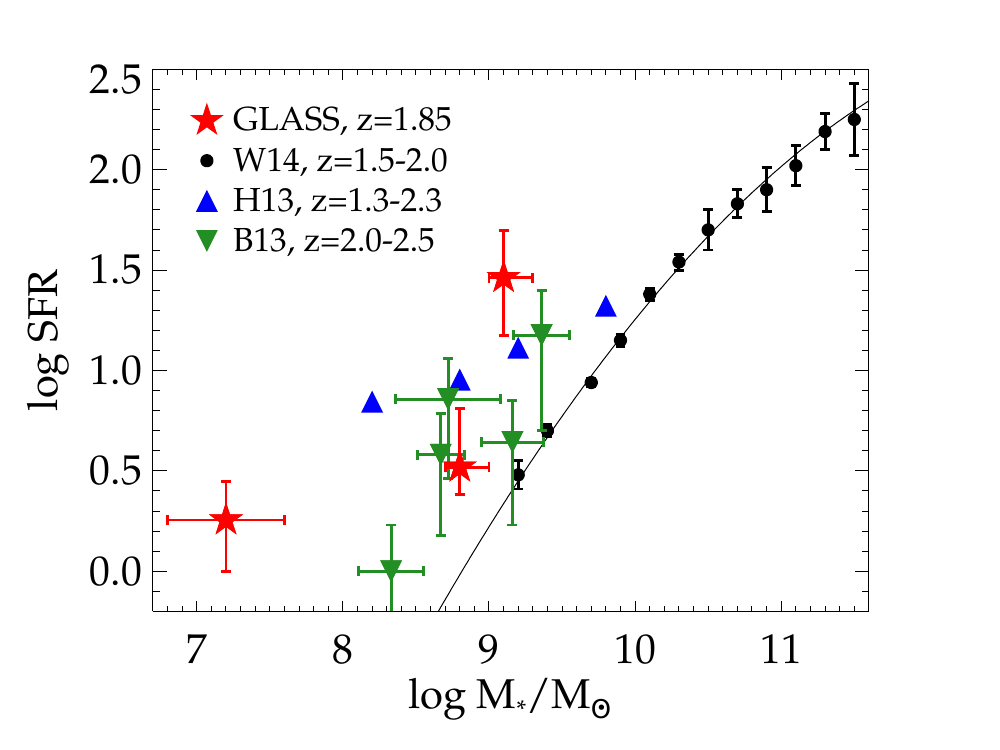} \\
\includegraphics[width=\columnwidth]{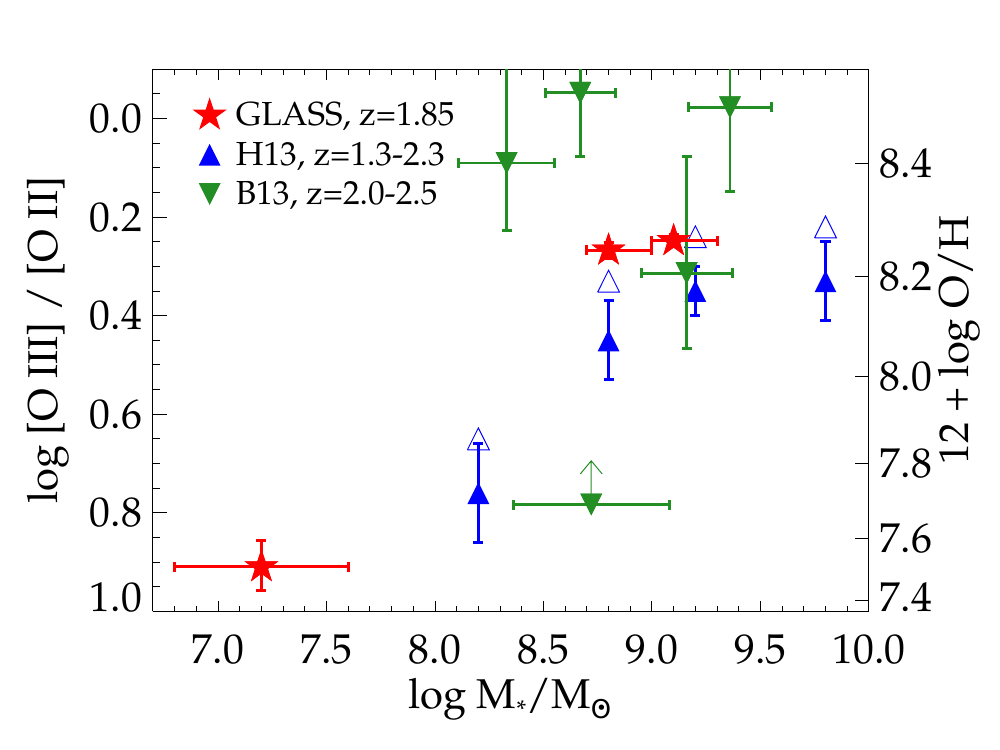}
\caption{
\label{fig:properties}
Global demographic properties of the GLASS arcs. For comparison we also show average values of larger samples at similar redshift from WISP \citep[][H13]{Henry2013} and 3D-HST \citep[][W14]{Whitaker2014} data as well as individual lensed galaxies reported by \citet[][B13]{Belli2013}. {\it Top:} SFR versus stellar mass. The arcs studied here have significantly higher specific SFR compared to the "main sequence" at $z\simeq1.8$ (shown by the W14 data), consistent with expectations for excess star formation triggered by gravitational interactions among these systems. {\it Bottom:} comparison of emission line ratios versus stellar mass. Open symbols show the effect of dust attenuation measured by \cite{Henry2013}; all other data are uncorrected for reddening. The arcs studied here have typical \Oiii$/$\Oii\ ratios for their redshift, stellar mass, and SFR. The right axis shows equivalent metallicity corresponding to \Oiii$/$\Oii\ for the \cite{Maiolino2008} calibration used in this work. The GLASS data are consistent with a steep mass-metallicity relation as found by \cite{Henry2013}, and notably extend to an order of magnitude lower in stellar mass compared to previous studies at $z\simeq2$.}
\end{figure}

In this section we examine the GLASS sample studied here in the context of the general $z\simeq1.8$ galaxy population, as a prelude to discussing the implications of our results for galaxy evolution. Figure~\ref{fig:properties} summarizes the demographic properties compared to larger published samples at similar redshift. Comparison samples are carefully chosen to minimize systematic effects: the narrow range of redshift mitigates any evolutionary trends, all data points are calculated with the same IMF and stellar population templates, and all stellar masses account for the contribution of emission lines to broad-band photometry.
Ignoring this last effect would result in higher stellar masses by $\simeq0.15$ dex for arcs 4 and 14 and $\simeq0.6$ dex for arc 3, in excellent agreement with the mass-dependent estimates by \cite{Whitaker2014}. In lieu of metallicity, we use \Oiii$/$\Oii\ flux ratios for a clear comparison of observable properties. While this ratio is sensitive to metallicity, ionization parameter, and reddening, all of these quantities are intrinsically correlated and vary monotonically with \Oiii$/$\Oii\ \citep[e.g.,][]{Maiolino2008,Dopita2013}. Furthermore, this ratio offers the lowest statistical uncertainty and the greatest sensitivity to metallicity of the available strong-line diagnostics in the sub-solar regime. We can therefore directly compare {\em relative} oxygen abundances between different samples, noting that secondary parameter dependences result in $\simeq0.25$ dex RMS scatter between metallicity and \Oiii$/$\Oii.

In comparison to the overall population, the galaxies studied here have SFRs which are a factor of $\geq3\times$ above the "main sequence" locus of SFR and M$_*$ (Figure~\ref{fig:properties}). Such high relative SFRs are often induced by gravitational interactions, and indeed \cite{Stott2013} show that $\sim50$\% of sources at this SFR, M$_*$, and $z$ are merger events. The metallicities inferred from \Oiii$/$\Oii\ are consistent with comparison samples of similar SFR and M$_*$, as are the nebular extinction values. The GLASS data are also consistent with a steep mass-metallicity relation as found by \cite{Henry2013}, although the scatter in Figure~\ref{fig:properties} is large. To summarize, the global properties are typical of low-mass starbursts at $z\simeq2$ which are frequently associated with galaxy interactions.

\subsection{Metallicity gradient evolution}

\begin{figure}
\includegraphics[width=0.45\textwidth]{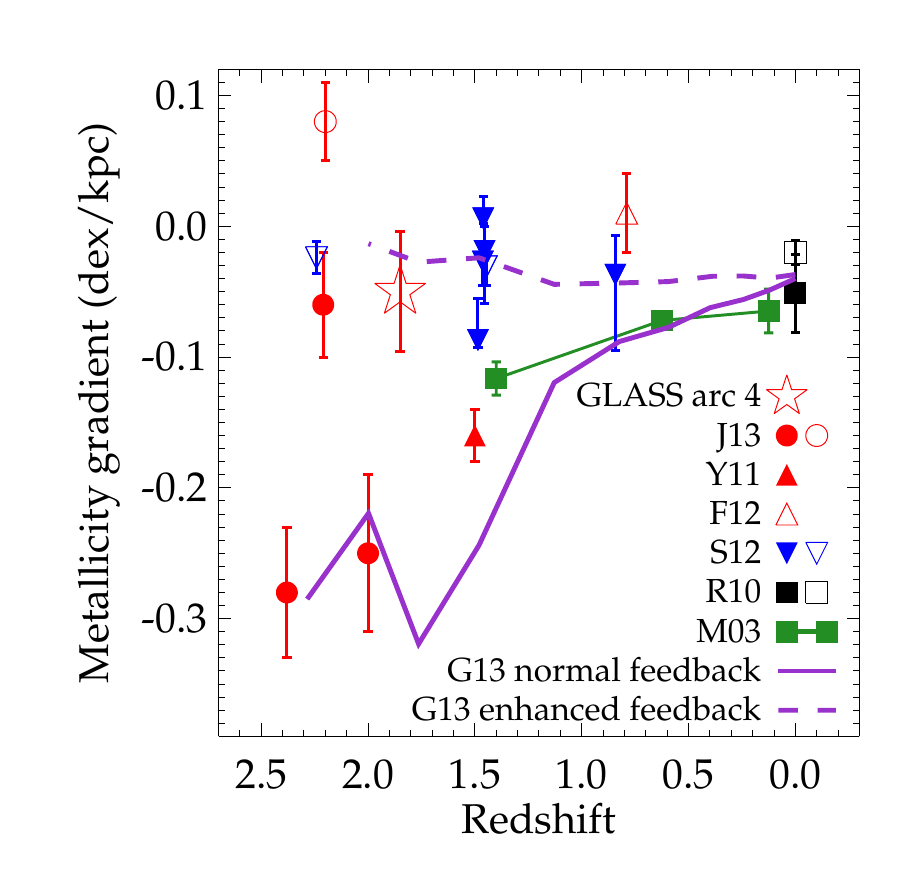}
\caption{
\label{fig:evolution}
Evolution of metallicity gradients with redshift. We compare the gradient of arc 4 from this work with published measurements at high redshift including other lensed galaxies (\citealt{Jones2013}, J13; \citealt{Yuan2011}, Y11; \citealt{Frye2012}, F12), non-lensed galaxies observed with adaptive optics \citep[][S12]{Swinbank2012}, an average of local gradients reported by \citet[][R10]{Rupke2010b}, and the Milky Way's metallicity gradient evolution measured from planetary nebulae \citep[][M03]{Maciel2003}. Solid and hollow symbols denote isolated disks and interacting (or merging) galaxies, respectively, to show that interacting galaxies have flatter gradients (closer to zero) on average in the lensed and $z=0$ samples. We additionally show results of two different feedback schemes in otherwise identical simulations, described in \cite[][G13; the galaxy shown is g15784]{Gibson2013}. The standard feedback results are similar to the Milky Way and isolated lensed galaxies, while enhanced feedback leads to shallower gradients at high redshifts and is good agreement with the measurement of arc 4 in this work.
}
\end{figure}

We now turn to a discussion of the metallicity gradient of arc 4 and its evolution with redshift. Figure~\ref{fig:evolution} compares the GLASS data with similar measurements as a function of redshift. Arcs 3 and 14 are not included because their gradients are poorly constrained, due to their small physical sizes and lower signal to noise ratio. In all three cases the gradients are consistent with zero at the $1\sigma$ level.
We restrict the comparison sample to measurements with $\lesssim1$ kpc resolution, noting that even kpc resolution results in artificially flat inferred gradients for typical high redshift galaxies \citep[e.g.,][]{Yuan2013,Stott2014}. We show original published values for all comparison data with the caveat that they are derived using different strong-line calibrations; this has minimal effect on our conclusions since metallicity gradients are relatively insensitive to the choice of calibration \citep{Jones2013}. The lensed $z=0.8$ source described by \cite{Frye2012} is also included with a previously unpublished gradient slope of $0.01\pm0.03$ dex kpc$^{-1}$ (B. Frye, private communication). However none of the comparison data include scatter in metallicity in the formal uncertainty, hence the error bar on arc 4 is the most conservative of those shown in Figure~\ref{fig:evolution}.
Ideally we would construct a sample which corresponds to approximately the same galaxy population \citep[as was done in][]{Jones2013}, but we are presently limited by the available data at high redshift. The result is that arc 4 is expected to have $\sim0.4$ dex lower stellar masses at a given redshift than average sources in Figure~\ref{fig:evolution}, which are approximately Milky Way analogs \citep[based on abundance matching; e.g.,][]{Behroozi2013}. Nonetheless this provides a useful comparison.

Figure~\ref{fig:evolution} demonstrates that interacting galaxies (shown as open symbols) have flatter gradients than isolated counterparts. This is most evident in the lensed and $z=0$ samples which have the best spatial resolution. Here we include arc 4 as an interacting system on the basis of its proximity to arc 14 and its enhanced SFR. Measurements of nearby galaxies have shown dramatically flattened gradients in pairs with the same projected separation as arcs 4--14 \citep[][and references therein]{Rich2012} which further motivates this classification. The relatively shallow metallicity gradient of arc 4 is therefore fully consistent with expected gravitational interaction. The gradient of arc 14 is also consistent with a near-zero value expected in this scenario. Arc 3 may have a shallow gradient as well, although this is not necessarily expected given its larger projected distance.

An alternative cause for shallow gradients measured in the GLASS data is metal mixing by strong feedback. To illustrate this, Figure~\ref{fig:evolution} shows evolutionary tracks for two simulations with different sub-grid feedback prescriptions of an otherwise identical Milky Way-like galaxy discussed by \cite{Gibson2013}, and we now briefly summarize their findings. The "normal" feedback simulation \citep[MUGS;][]{Stinson2010} results in relatively steep metallicity gradients of 0.2--0.3 dex\,kpc$^{-1}$ at $z\simeq2$ which flatten at later times. The MaGICC simulation \citep{Brook2011} employs an enhanced feedback scheme, in which galactic outflows are more effective at removing metal-enriched ISM material which is re-accreted preferentially in outer regions. This mixing of the ISM results in shallower gradients of $<0.05$ dex\,kpc$^{-1}$ which become marginally steeper with time. Other groups have found essentially identical behavior with various feedback mechanisms which act to redistribute energy and ISM material throughout galactic disks \citep[e.g.,][]{Yang2012,Angles-Alcazar2014}. 
The enhanced feedback results in Figure~\ref{fig:evolution} are in excellent agreement with the shallow gradient measured for arc 4 as well as for arc 14. Arc 3 is expected to have a steeper negative gradient than shown by these models due to its lower mass \citep[e.g.,][]{Few2012}, but we are unable to distinguish between a shallow and steep negative gradient within the uncertainty.
The overall metallicity can in principle help to identify the cause of shallow gradients: rapidly recycled outflows should produce a signature of higher metallicity, whereas gravitational interactions should have an opposite effect \citep[e.g.,][]{Rich2012}. We note that \cite{Kulas2013} report evidence for rapid recycling in a protocluster at $z=2.3$ based on stacked spectra, and we might expect a similar observational signature in the overdense region studied here. However, the expected differences are small \citep[$\sim0.1$ dex;][]{Torrey2012,Kulas2013} compared to measurement uncertainties so we are unable to distinguish these scenarios without additional complementary data. In any case the shallow gradient in arc 4 suggests that metals are efficiently mixed throughout the ISM on scales of several kpc.

\subsection{A starbursting dwarf galaxy at $z\simeq2$}

One of the most striking aspects of Figure~\ref{fig:properties} is the low stellar mass of arc 3, M$_* = 1.5 \times 10^7 \Msun$. This is an order of magnitude below previous metallicity studies at $z\simeq2$ and extends into the regime of Milky Way dwarf satellites. Since such galaxies have rarely been studied directly at high redshift \citep{Christensen2012,Stark2014,Atek2014}, we briefly consider the properties of arc 3 in the context of recent cosmological simulations and local group analogs.

Both simulations and observations of nearby dwarf galaxies suggest that in the mass regime of arc 3, $\sim$50\% of its present-day stellar mass has already formed by $z=1.85$ \citep{Shen2014,Weisz2011}. In terms of mass it is therefore roughly analogous to the "Doc" simulation of \cite{Shen2014}, and to the Milky Way dwarf spheroidal Fornax \citep[e.g.,][]{Coleman2008}. This is supported by the overall oxygen abundance \oh\,$=7.75$ (equivalent to [O/H]~$=-0.94$) which is in excellent agreement with the stellar $\alpha$-element abundance distribution of Fornax \citep{Kirby2011}. Figure~\ref{fig:gradients} indicates a gas-phase metallicity gradient of $-0.29\pm0.25$ which is also compatible with Fornax's {\em present-day stellar} metallicity gradient \citep[$-0.50\pm0.10$ dex\,kpc$^{-1}$;][]{Hendricks2014}. However the uncertainty is large, and these two quantities need not be identical if there is significant radial growth or dynamical evolution of stellar orbits. Such a steep gradient would imply that any dynamical interaction in arc 3 is insufficient for strong radial mixing, and this is consistent with \cite{Rich2012} given that the nearest detected neighbor (arc 14) is separated by 150 kpc in projection.

We measure an instantaneous specific star formation rate of $\log{\rm SSFR (yr^{-1})} = -7.0 \pm 0.5$ for arc 3, i.e., a mass doubling time of only $10^{7\pm0.5}$ yr. However we caution that this is calculated assuming constant SFR over the previous $\sim20$ Myr, which is likely not the case: such a high value, as well as the high ratio of Balmer line flux to UV luminosity and the stellar population modeling described in Section~\ref{sec:mstar}, suggests a short duty cycle for the current burst of star formation.
Such intense short starbursts are predicted by several simulations of dwarf galaxies \citep[e.g.,][]{Governato2012,Zolotov2012,Brooks2014,Shen2014}, and may be triggered by interactions or by gradual accumulation of a large gas reservoir. Bursty star formation histories are of particular interest in light of observations that the central regions of local group dwarf galaxies are less dense than predicted for $\Lambda$CDM cosmology in the absence of baryonic feedback \citep[the "too big to fail" problem;][]{Boylan-Kolchin2011}. These simulations have found that repeated bursts of star formation and strong feedback in dwarf galaxies can affect the gravitational potential, transforming dark matter density profiles from cusps to cores, and have been postulated to reconcile $\Lambda$CDM with observations of local group substructure \citep[although see also][]{Garrison-Kimmel2013}.
The high SFR surface density of arc 3 is sufficient to drive outflows with high mass loading factors \citep{Newman2012} as required by these simulations. However we do not directly constrain the outflow properties, and direct measurements of outflowing gas are extremely challenging for the luminosity of arc 3. One means of making progress is to measure the mass-metallicity relation slope which is sensitive to the properties of stellar feedback. The data shown in Figure~\ref{fig:properties} are consistent with feedback in the form of strong energy-driven galactic winds \citep{Henry2013}, but of course a larger sample will be required to better constrain the mass-metallicity relation at M$_*\lesssim10^8 \Msun$. We expect to find several dozen such low-mass sources in the full GLASS survey, which will represent the first statistical constraint of chemical enrichment and feedback at high redshift in the mass regime where galaxies are in tension with $\Lambda$CDM. The present data for arc 3 are generally consistent with simulations in which bursty star formation and strong feedback significantly lower the central density of dwarf galaxies.

\section{Conclusions}\label{sec:conclusions}

This work presents a case study of galaxy evolution in a dense environment at $z=1.85$ from the GLASS survey, utilizing the spatial resolution of \hst\ aided by gravitational lensing. We study three galaxies in close physical proximity (50--200 kpc projected separations) and measure their spatially resolved metallicities from \hst\ grism spectra. Our main results are as follows:
\begin{itemize}
\item All three galaxies appear to be affected by interaction-induced gas flows, evidenced by their close proximity and specific star formation rates which are $\geq3\times$ above the median value for their M$_*$ and redshift.
\item We measure a precise metallicity gradient for one galaxy and find a shallow slope compared to other sources at similar redshift, as expected from gravitational interactions. The gradients are also consistent with strong stellar feedback resulting in higher metallicity of accreted (recycled) gas, although we find no evidence of enhanced metallicity which should result from this effect.
\item The stellar mass range extends an order of magnitude below previous metallicity surveys at $z\simeq2$. The data are consistent with a steep mass-metallicity relation found in previous work, and the full GLASS survey should yield a sufficient sample to measure metallicity trends in the range M$_* \simeq10^7-10^8 \Msun$.
\item We present the first spatially resolved spectrophotometric analysis of a bona-fide dwarf galaxy at $z\simeq2$ with stellar mass M$_*=10^{7.2\pm0.4}\, \Msun$. It is analogous to Fornax in terms of its stellar mass, metallicity, and metallicity gradient. Its extremely high SSFR indicates a violent starburst, likely with a low duty cycle, consistent with the hypothesis that successive starbursts with strong feedback are responsible for flattening the central density profiles of dwarf galaxies.
\end{itemize}

Measuring accurate metallicity gradients at high redshift is a significant observational challenge which, when overcome, is expected to elucidate various aspects of the galaxy formation process. Observing lensed galaxies at $\lesssim0\farcs2$ resolution is the best way to obtain the requisite spatial sampling \citep{Yuan2013}. Previously only 5 such gradient measurements have been reported in the literature \citep{Jones2010,Jones2013,Yuan2011}, although we note that improvements to the Keck/OSIRIS spectrograph and adaptive optics system have enabled recent observations of a significantly larger sample (Leethochawalit et al. in prep). Our analysis of \hst\ data provides further support for a scenario in which gravitational interactions and mergers cause a temporary flattening of gradients, while isolated galaxies at high redshift have steep negative metallicity gradients consistent with standard feedback models. More importantly, our results constitute a proof-of-concept that \hst\ grism spectra obtained with the GLASS survey yield precise gradient slope measurements with conservative uncertainty $\simeq0.05$ dex\,kpc$^{-1}$ in good cases. This is enabled by the combination of broad wavelength coverage and high spatial resolution aided by gravitational lensing. We note that similar previous work has shown the benefits of this approach, including several studies of lensed galaxies observed with \hst\ grisms \citep{Brammer2012, Frye2012, Whitaker2014b}.
This paper presents the complete methodology developed for such measurements, including a significant improvement over previous lensing studies by combining data from multiple strongly lensed images (Section~\ref{sec:model}; Wang et al. in prep). From visual inspection of the current GLASS data, we identify $\simeq5-20$ galaxies per cluster at $z=1-3$ with multiple bright emission lines suitable for resolved metallicity analysis. We expect the full survey to yield high-quality gradient measurements (comparable to arc 4) for approximately 20 of these galaxies. (This metallicity gradient sample will most likely be composed of relatively massive galaxies with M$_* \gtrsim 10^9 \, \Msun$, while at the same time we expect to characterize the properties of a larger sample of low-mass galaxies with modest spatial information).
We anticipate that such a large sample will conclusively establish whether metallicity gradient slopes are steeper or shallower at $z\simeq2$ compared to local descendants, and more generally how metallicity gradients evolve over time. This will have significant implications for sub-grid feedback prescriptions used in cosmological simulations \citep[e.g.,][]{Gibson2013,Angles-Alcazar2014}, which in turn inform our understanding of baryon cycling and chemical enrichment of the circum- and inter-galactic media, and the role of this baryon cycle in galaxy formation.

\acknowledgements

We thank R. Maiolino for providing the metallicity calibrations used in this work, and B. Gibson for providing metallicity gradient evolution tracks from simulations. We thank the anonymous referee for a constructive report which improved the clarity of this paper. T. Johnson and K. Sharon are acknowledged for providing the map of the phase angle.
TAJ acknowledges support from the Southern California Center for Galaxy Evolution through a CGE Fellowship.
We acknowledge support from NASA through grant HST-13459. TT acknowledges support by the Packard Foundation in the form of a Packard Research Fellowship and thanks the American Academy in Rome and the Observatory of Monteporzio Catone for their generous hospitality.
This paper is based on observations made with the NASA/ESA Hubble Space Telescope, and utilizes gravitational lensing models produced by PIs Brada{\v c}, Ebeling, Merten \& Zitrin, Sharon, and Williams funded as part of the HST Frontier Fields program conducted by STScI. STScI is operated by the Association of Universities for Research in Astronomy, Inc. under NASA contract NAS 5-26555. The lens models were obtained from the Mikulski Archive for Space Telescopes (MAST).


\begin{thebibliography}{43}
\expandafter\ifx\csname natexlab\endcsname\relax\def\natexlab#1{#1}\fi

\bibitem[{Allende Prieto} {et~al.}(2001)]{Allende2001} Allende Prieto, C., Lambert, D.~L., \& Asplund, M.\ 2001, \apjl, 556, L63

\bibitem[Angl{\'e}s-Alc{\'a}zar et al.(2014)]{Angles-Alcazar2014} Angl{\'e}s-Alc{\'a}zar, D., Dav{\'e}, R., {\"O}zel, F., \& Oppenheimer, B.~D.\ 2014, \apj, 782, 84

\bibitem[Atek et al.(2014)]{Atek2014} Atek, H., et al.\ 2014, \apj, 789, 96

\bibitem[Behroozi et al.(2013)]{Behroozi2013} Behroozi, P.~S., Wechsler, R.~H., \& Conroy, C.\ 2013, \apj, 770, 57

\bibitem[Belli et al.(2013)]{Belli2013} Belli, S., Jones, T., Ellis, R.~S., \& Richard, J.\ 2013, \apj, 772, 141

\bibitem[Bertin \& Arnouts(1996)]{Bertin1996} Bertin, E., \& Arnouts, S.\ 1996, \aaps, 117, 393

\bibitem[Boylan-Kolchin et al.(2011)]{Boylan-Kolchin2011} Boylan-Kolchin, M., Bullock, J.~S., \& Kaplinghat, M.\ 2011, \mnras, 415, L40

\bibitem[Brammer et al.(2012)]{Brammer2012} Brammer, G.~B., et al.\ 2012, \apjl, 758, LL17 

\bibitem[Bresolin et al.(2012)]{Bresolin2012} Bresolin, F., Kennicutt, R.~C., \& Ryan-Weber, E.\ 2012, \apj, 750, 122

\bibitem[Brook et al.(2011)]{Brook2011} Brook, C.~B., et al.\ 2011, \mnras,
415, 1051

\bibitem[Brooks \& Zolotov(2014)]{Brooks2014} Brooks, A.~M., \& Zolotov, A.\ 2014, \apj, 786, 87

\bibitem[Bruzual \& Charlot(2003)]{Bruzual2003} Bruzual, G., \& Charlot, S.\ 2003, \mnras, 344, 1000

\bibitem[Cardelli, Clayton, \& Mathis(1989)]{Cardelli1989} Cardelli, J.~A., Clayton, G.~C., \& Mathis, J.~S.\ 1989, \apj, 345, 245

\bibitem[Chabrier(2003)]{Chabrier2003} Chabrier, G.\ 2003, \apjl, 586, L133

\bibitem[Chiappini et al.(2001)]{Chiappini2001} Chiappini, C., Matteucci, F., \& Romano, D.\ 2001, \apj, 554, 1044

\bibitem[Christensen et al.(2012)]{Christensen2012} Christensen, L., et
al.\ 2012, \mnras, 427, 1953

\bibitem[Coleman \& de Jong(2008)]{Coleman2008} Coleman, M.~G., \& de Jong, J.~T.~A.\ 2008, \apj, 685, 933

\bibitem[Cresci et al.(2010)]{Cresci2010} Cresci, G., Mannucci, F., Maiolino, R., et al.\ 2010, \nat, 467, 811

\bibitem[Dav{\'e} et al.(2011)]{Dave2011} Dav{\'e}, R., Finlator, K., \& Oppenheimer, B.~D.\ 2011, \mnras, 416, 1354

\bibitem[Dekel et al.(2009)]{Dekel2009} Dekel, A., Sari, R., \& Ceverino, D.\ 2009, \apj, 703, 785

\bibitem[Dopita et al.(2013)]{Dopita2013} Dopita, M.~A., Sutherland, R.~S.,
Nicholls, D.~C., Kewley, L.~J., \& Vogt, F.~P.~A.\ 2013, \apjs, 208, 10

\bibitem[Faucher-Gigu{\`e}re et al.(2011)]{Faucher-Giguere2011} Faucher-Gigu{\`e}re, C.-A., Kere{\v s}, D., \& Ma, C.-P.\ 2011, \mnras, 417, 2982

\bibitem[Few et al.(2012)]{Few2012} Few, C.~G., Gibson, B.~K., Courty, S., et al.\ 2012, \aap, 547, A63 

\bibitem[Frye et al.(2012)]{Frye2012} Frye, B.~L., et al.\ 2012, \apj, 754,  17 

\bibitem[Fu et al.(2009)]{Fu2009} Fu, J., Hou, J.~L., Yin, J., \& Chang, R.~X.\ 2009, \apj, 696, 668

\bibitem[Garrison-Kimmel et al.(2013)]{Garrison-Kimmel2013} Garrison-Kimmel, S., Rocha, M., Boylan-Kolchin, M., Bullock, J.~S., \& Lally, J.\ 2013, \mnras, 433, 3539

\bibitem[Gibson et al.(2013)]{Gibson2013} Gibson, B.~K., Pilkington, K., Brook, C.~B., Stinson, G.~S., \& Bailin, J.\ 2013, \aap, 554, A47

\bibitem[Governato et al.(2012)]{Governato2012} Governato, F., et al.\
2012, \mnras, 422, 1231

\bibitem[Heckman(2001)]{Heckman2001} Heckman, T.~M.\ 2001, Gas and Galaxy
Evolution, 240, 345

\bibitem[Hendricks et al.(2014)]{Hendricks2014} Hendricks, B., Koch, A., 
Walker, M., Johnson, C.~I., Pe{\~n}arrubia, J., \& Gilmore, G.\ 2014, \aap, 572, AA82 

\bibitem[Henry et al.(2013)]{Henry2013} Henry, A., et al.\ 2013, \apjl,
776, L27

\bibitem[Hummer \& Storey(1987)]{Hummer1987} Hummer, D.~G., \& Storey, P.~J.\ 1987, \mnras, 224, 801

\bibitem[Johnson et al.(2014)]{Johnson2014} Johnson, T.~L., Sharon, K., 
Bayliss, M.~B., Gladders, M.~D., Coe, D., \& Ebeling, H.\ 2014, \apj, 797, 48 

\bibitem[Jones et al.(2010)]{Jones2010} Jones, T., Ellis, R., Jullo, E., \& Richard, J.\ 2010, \apjl, 725, L176

\bibitem[Jones et al.(2013)]{Jones2013} Jones, T., Ellis, R.~S., Richard, J., \& Jullo, E.\ 2013, \apj, 765, 48

\bibitem[Kennicutt(1998)]{Kennicutt1998} Kennicutt, R.~C., Jr.\ 1998,
\araa, 36, 189

\bibitem[Kere{\v s} et al.(2009)]{Keres2009} Kere{\v s}, D., Katz, N.,
Fardal, M., Dav{\'e}, R., \& Weinberg, D.~H.\ 2009, \mnras, 395, 160

\bibitem[Kirby et al.(2011)]{Kirby2011} Kirby, E.~N., Cohen, J.~G., Smith,
G.~H., Majewski, S.~R., Sohn, S.~T., \& Guhathakurta, P.\ 2011, \apj, 727, 79

\bibitem[Kriek et al.(2009)]{Kriek2009} Kriek, M., van Dokkum, P.~G., Labb{\'e}, I., Franx, M., Illingworth, G.~D., Marchesini, D., \& Quadri, R.~F.\ 2009, \apj, 700, 221

\bibitem[Kulas et al.(2013)]{Kulas2013} Kulas, K.~R., et al.\ 2013, \apj,
774, 130

\bibitem[Lamareille(2010)]{Lamareille2010} Lamareille, F.\ 2010, \aap, 509,
A53

\bibitem[Lequeux et al.(1979)]{Lequeux1979} Lequeux, J., Peimbert, M.,
Rayo, J.~F., Serrano, A., \& Torres-Peimbert, S.\ 1979, \aap, 80, 155

\bibitem[Limousin et al.(2012)]{Limousin2012} Limousin, M., et al.\ 2012,
\aap, 544, A71

\bibitem[Maciel et al.(2003)]{Maciel2003} Maciel, W.~J., Costa, R.~D.~D., \& Uchida, M.~M.~M.\ 2003, \aap, 397, 667

\bibitem[Magrini et al.(2007)]{Magrini2007} Magrini, L., Corbelli, E., \& Galli, D.\ 2007, \aap, 470, 843

\bibitem[Maiolino et al.(2008)]{Maiolino2008} Maiolino, R., et al.\ 2008,
\aap, 488, 463

\bibitem[Nelson et al.(2012)]{Nelson2012} Nelson, E.~J., et al.\ 2012,
\apjl, 747, L28

\bibitem[Newman et al.(2012)]{Newman2012} Newman, S.~F., et al.\ 2012,
\apj, 761, 43

\bibitem[Newman et al.(2014)]{Newman2014} Newman, S.~F., et al.\ 2014,
\apj, 781, 21

\bibitem[Pettini \& Pagel(2004)]{Pettini2004} Pettini, M., \& Pagel, B.~E.~J.\ 2004, \mnras, 348, L59

\bibitem[Pilkington et al.(2012)]{Pilkington2012} Pilkington, K., et al.\
2012, \aap, 540, A56

\bibitem[Planck Collaboration (2014)]{Planck2013} Planck Collaboration: Ade, P. A. R., et al.\ 2014, \aap, 571, AA16 

\bibitem[Postman et al.(2012)]{Postman2012} Postman, M., et al.\ 2012,
\apjs, 199, 25

\bibitem[Prantzos \& Boissier(2000)]{Prantzos2000} Prantzos, N., \& Boissier, S.\ 2000, \mnras, 313, 338

\bibitem[Queyrel et al.(2012)]{Queyrel2012} Queyrel, J., Contini, T., Kissler-Patig, M., et al.\ 2012, \aap, 539, A93

\bibitem[Rich et al.(2012)]{Rich2012} Rich, J.~A., Torrey, P., Kewley,
L.~J., Dopita, M.~A., \& Rupke, D.~S.~N.\ 2012, \apj, 753, 5

\bibitem[Rupke et al.(2010a)]{Rupke2010a} Rupke, D.~S.~N., Kewley, L.~J., \& Barnes, J.~E.\ 2010, \apjl, 710, L156

\bibitem[Rupke et al.(2010b)]{Rupke2010b} Rupke, D.~S.~N., Kewley, L.~J., \& Chien, L.-H.\ 2010, \apj, 723, 1255

\bibitem[S{\'a}nchez et al.(2014)]{Sanchez2014} S{\'a}nchez, S.~F., et
al.\ 2014, \aap, 563, A49

\bibitem[Schmidt et al.(2014)]{Schmidt2014} Schmidt, K.~B., et al.\ 2014,
\apjl, 782, L36

\bibitem[Shapley et al.(2014)]{Shapley2014} Shapley, A.~E., Reddy, N.~A., Kriek, M., et al.\ 2014, arXiv:1409.7071 

\bibitem[Shen et al.(2014)]{Shen2014} Shen, S., Madau, P., Conroy, C., Governato, F., \& Mayer, L.\ 2014, \apj, 792, 99

\bibitem[Stark et al.(2014)]{Stark2014} Stark, D.~P., et al.\ 2014, \mnras, 
445, 3200 

\bibitem[Steidel et al.(2014)]{Steidel2014} Steidel, C.~C., et al.\ 2014, 
\apj, 795, 165 

\bibitem[Stinson et al.(2010)]{Stinson2010} Stinson, G.~S., Bailin, J., Couchman, H., Wadsley, J., Shen, S., Nickerson, S., Brook, C., \& Quinn, T.\ 2010, \mnras, 408, 812

\bibitem[Stott et al.(2013)]{Stott2013} Stott, J.~P., Sobral, D., Smail,
I., Bower, R., Best, P.~N., \& Geach, J.~E.\ 2013, \mnras, 430, 1158

\bibitem[Stott et al.(2014)]{Stott2014} Stott, J.~P., et al.\ 2014, \mnras,
443, 2695

\bibitem[Swinbank et al.(2012)]{Swinbank2012} Swinbank, A.~M., Sobral, D.,
Smail, I., Geach, J.~E., Best, P.~N., McCarthy, I.~G., Crain, R.~A., \& Theuns, T.\ 2012, \mnras, 426, 935

\bibitem[Torrey et al.(2012)]{Torrey2012} Torrey, P., Cox, T.~J., Kewley, L., \& Hernquist, L.\ 2012, \apj, 746, 108

\bibitem[Tremonti et al.(2004)]{Tremonti2004} Tremonti, C.~A., et al.\
2004, \apj, 613, 898

\bibitem[Troncoso et al.(2014)]{Troncoso2014} Troncoso, P., et al.\ 2014,
\aap, 563, A58

\bibitem[Trump et al.(2011)]{Trump2011} Trump, J.~R., et al.\ 2011, \apj,
743, 144

\bibitem[Vila-Costas \& Edmunds(1992)]{Vila-Costas1992} Vila-Costas, M.~B., \& Edmunds, M.~G.\ 1992, \mnras, 259, 121

\bibitem[Weisz et al.(2011)]{Weisz2011} Weisz, D.~R., et al.\ 2011, \apj,
739, 5

\bibitem[Werk et al.(2011)]{Werk2011} Werk, J.~K., Putman, M.~E., Meurer, G.~R., \& Santiago-Figueroa, N.\ 2011, \apj, 735, 71

\bibitem[Whitaker et al.(2014)]{Whitaker2014b} Whitaker, K.~E., Rigby, 
J.~R., Brammer, G.~B., Gladders, M.~D., Sharon, K., Teng, S.~H., 
\& Wuyts, E.\ 2014, \apj, 790, 143 

\bibitem[Whitaker et al.(2014)]{Whitaker2014} Whitaker, K.~E., et al.\ 
2014, \apj, 795, 104 

\bibitem[Wright et al.(2010)]{Wright2010} Wright, S.~A., Larkin, J.~E., Graham, J.~R., \& Ma, C.-P.\ 2010, \apj, 711, 1291

\bibitem[Yang \& Krumholz(2012)]{Yang2012} Yang, C.-C., \& Krumholz, M.\ 2012, \apj, 758, 48

\bibitem[Yuan et al.(2011)]{Yuan2011} Yuan, T.-T., Kewley, L.~J., Swinbank,
A.~M., Richard, J., \& Livermore, R.~C.\ 2011, \apjl, 732, L14

\bibitem[Yuan et al.(2012)]{Yuan2012} Yuan, T.-T., Kewley, L.~J., Swinbank, A.~M., \& Richard, J.\ 2012, \apj, 759, 66

\bibitem[Yuan et al.(2013)]{Yuan2013} Yuan, T.-T., Kewley, L.~J., \& Rich, J.\ 2013, \apj, 767, 106

\bibitem[Zolotov et al.(2012)]{Zolotov2012} Zolotov, A., et al.\ 2012,
\apj, 761, 71


\end{thebibliography}
\end{document}